\documentclass[preprintnumbers,showpacs,showkeys,twocolumn,prb]{revtex4}

\usepackage{graphicx}
\usepackage{epsf}
\usepackage{dcolumn}
\usepackage{bm}

\begin{document}

\preprint{LA-UR:02-3843}

\title{Limits on phonon information extracted from neutron pair-density 
functions}

\author{Matthias J. Graf, Il-Kyoung Jeong, Daniel L. Starr, R.~H. Heffner}
\affiliation{Los Alamos National Laboratory, Los Alamos, New Mexico 87545}

\date{\today}

\begin{abstract}
We explore the possibility of extracting information about lattice 
dynamics in simple crystal structures from the neutron pair-density function 
(PDF) through inverse data analysis. Contrary to the claims by 
Dimitrov, Louca, and R\"oder [Phys.~Rev.~B {\bf 60}, 6204 (1999)],
and in agreement with recent work by
Reichardt and Pintschovius [Phys.~Rev.~B {\bf 63}, 174302 (2001)],
we find that the PDF alone is not sufficient for constructing accurate
phonon dispersions in the entire Brillouin zone in systems with complex
lattice dynamics.
However, our numerical simulations show that for monatomic 
{\it fcc} and {\it bcc} crystal structures it is, in principle, 
possible to obtain phonon moments of complex metals as well as phonon 
frequencies of simple metals within a few percent accuracy.
\end{abstract}


\maketitle


\section{Introduction}

Our goal in this work is to determine whether it is possible to obtain 
high-quality information about the phonon dispersion and lattice dynamics 
over the entire Brillouin zone (BZ) without having to measure the phonon 
dispersion in single crystals. This is an important question for materials 
where no single crystals are available, or when performing measurements at 
high pressure and high temperature.
This work was motivated by the claim of Dimitrov and co-workers 
\cite{dimitrov99} that it is possible to extract accurate phonon dispersions 
by properly modeling the phonon system and applying an inverse data 
analysis technique to the {\it measured} neutron pair-density function (PDF). 
In principle, the PDF contains all the lattice dynamical information,
although most of it is lost after the integration over dynamical and 
directional degrees of freedom is performed.
Very recently, Reichardt and Pintschovius \cite{pintschovius01} seriously 
questioned the results of Dimitrov {\it et al.},\cite{dimitrov99}
and concluded that the PDF is rather insensitive to the precise shape of the 
dispersion curves, and, therefore, that there is no hope for extracting 
lattice vibrations from experimental PDF measurements.
Similar concerns were raised by Mellerg{\aa}rd and McGreevy.\cite{mellergard00}

Experimental PDF data obtained from neutron powder diffraction require various 
corrections to the raw data, which introduce additional errors in the analysis.
Thus, in this paper we studied \textit{synthetic} data sets to avoid
any ambiguity in the analysis. 
We tested and utilized standard reverse Monte Carlo\cite{MonteCarlo}
and Levenberg-Marquardt\cite{recipes}
methods for the inverse analysis of the PDF spectra to determine
how much lattice dynamical information can be recovered.
To do that, we generated synthetic PDF data sets for various monatomic
{\it fcc} and {\it bcc} crystal structures from published tables of 
generalized Born-von K\'arm\'an (BvK) force constants.\cite{Landolt}
We used a BvK model to describe the lattice dynamics since it is easy to 
implement and is well documented in the literature, although it is not
the best-suited phonon model for metals.
This gave us full control when testing the robustness and accuracy of the
inverse analysis methods being used.
We emphasize that we  wished to determine whether a unique relationship 
exists between the phonon dispersion and the corresponding PDF of 
the system, {\it not} a unique relationship between the force constants
of the phonon model and its dispersion or PDF spectrum.

Our study of the {inverse} (indirect) problem of extracting phonons
from a given PDF spectrum complements the previous studies by 
Reichardt and Pintschovius,\cite{pintschovius01} who studied the 
{forward} (direct) problem of obtaining a unique PDF spectrum from 
very different phonon dispersion curves.

\section{Analysis}

We calculated the PDF spectrum by Fourier transforming the powder-averaged,
coherent static structure factor, $S(q)$,\cite{squires,lovesey,toby92}
\begin{equation}
\varrho(r) = \varrho_0 + \frac{1}{2 \pi^2 r}
\int_{0}^{Q} dq \, q \sin(qr) \big[ S(q) - 1 \big]
\,,
\end{equation}
with atomic number density $\varrho_0$.
For simplicity, we neglected multiphonon processes ({\it i.e.}, two-phonon
and higher-order processes) in the computation of the diffuse scattering part 
of $S(q)$.\cite{multiphonon} Thus, we assumed that the one-phonon scattering 
process is the dominant inelastic scattering process in the range of usually
\textit{measured} scattering vectors, $q < Q \sim 35-40 {\rm \AA}^{-1}$. 
However, as was pointed out earlier, \cite{chung97,earlier,pintschovius01} 
one needs to be more careful about multiphonon contributions when comparing 
with actual experiments.

We fit the synthetic PDF data sets $\varrho_{\rm synth}(r_i)$ with respect
to the force constants of the model by minimizing the function
\begin{equation}\label{chi2pdf}
\chi^2_{\rm pdf} = \frac{1}{N-F} \sum_{i=1}^{N}
 \frac{\big[ \varrho_{\rm synth}(r_i) - \varrho(r_i) 
       \big]^2}{\sigma^2(r_i)} \,,
\end{equation}
where $F$ is the number of force constants being fitted, and $N \gg F$.
The number of spatial points is of the order of $N \sim 1000$, and
$\sigma(r_i) = \varepsilon_\sigma \varrho(r_*) r_*/r_i$ 
is an error estimate for $\varrho_{\rm synth}(r_i)$.\cite{toby92}
We use a relative error $\varepsilon_\sigma\approx 0.03$ at the first 
PDF peak maximum at distance $r_*$, which is typical for experiments 
with good neutron counting statistics and proper background corrections.
Values as low as $\varepsilon_\sigma\approx 0.01$ are feasible in 
high-precision diffraction experiments. Since $\varepsilon_\sigma$ enters in
Eq.~\ref{chi2pdf} only as an overall scaling factor, none of our results
depend on the absolute value of $\varepsilon_\sigma$.
The sums run from just below the first peak through $r_N=1{\rm nm}$. 
Extending the sum to $r_N = 2{\rm nm}$ does not lead to any significant 
changes in our results.
The model PDF $\varrho(r_i)$ depends implicitly on the phonon dispersion and,
thus, on the fitted force constants.

One starts the loop of the fitting procedure with a (small) set of plausible
force constants and calculates the phonon frequencies $f({\bf k}s)$
and eigenvectors ${\bf e}({\bf k}s)$ on a fine mesh in the BZ.
The computation of the static structure factor in the one-phonon approximation
is straightforward, once the frequencies and eigenvectors are known.
Finally, one computes the model PDF by convoluting the static structure 
factor with the instrumental resolution function of the 
diffractometer \cite{GSAS} 
and compares it with the synthetic PDF. An update of the force constants 
follows. These steps are repeated until a termination criterion is met, 
either in the implemented reverse Monte Carlo or Levenberg-Marquardt method.
Since the Monte Carlo method is computationally intensive and slow compared
to a generalized nonlinear least-squares minimizer, we tested the reverse
Monte Carlo method only for models with a small set of fitting parameters.

Next we checked the quality of the resulting phonon dispersions by comparing
the second phonon moment $f_2$ and the phonon dispersion 
$f({\bf k}s)$ with the synthetic phonon data.
The second moment of a monatomic crystal is defined by\cite{Landolt}
\begin{equation}
f_2  =  \left( \frac{5}{9 {N_{\rm BZ}}}
	\sum_{s=1}^{3} \sum_{{\bf k} \in {\rm BZ}}  f^2({\bf k}s)
        \right)^{1/2} \,,
\end{equation}
with wave vector ${\bf k}$, phonon branch index $s$, 
and ${N_{\rm BZ}}$ {\bf k}-points in the summation over the Brillouin zone.

In order to quantify the goodness of the phonon dispersion we introduce 
the merit function
\begin{equation}
\chi^2_{\rm phon} = \frac{1}{N_p - F} \sum_{i=1}^{N_p}
  \frac{\big[ f_{\rm synth}({\bf k}_i s_i) - f({\bf k}_i s_i) 
        \big]^2}{\sigma_p^2({\bf k}_i s_i)} \,,
\end{equation}
where $N_p$ is the number of phonon frequencies. Their estimated errors
$\sigma_p$ are taken to be two percent of the frequency, but at least 
$0.04 {\rm THz}$.
These are typical error estimates reported for neutron triple-axis 
spectroscopy experiments.

\section{Results}

We studied in detail elemental Ni, Ag, Al, Ce and Pb ({\it fcc}), as well as 
Fe and Nb ({\it bcc}). Since we obtained very similar results for Ni and Ag,
we will not discuss Ag separately.
The results for {\it bcc} structures show   
very similar behavior to the ones with {\it fcc} structure. 
Ni has a simple phonon dispersion typical of monatomic {\it fcc} crystals 
that can be described very well by including only the first few 
nearest-neighbor (NN) shells of interatomic force constants in a BvK model. 
On the other hand, Pb shows a complex dispersion (even when neglecting 
Kohn anomalies) that requires long-range forces up through 8NN shells in
a BvK model. 
Similarly, Fe is the prototype of a monatomic {\it bcc} crystal with a
simple phonon dispersion, whereas Nb possesses a very complex dispersion.

\begin{figure}[floatfix]
\noindent
\begin{center}
\epsfysize=80mm
\rotatebox{-90}{
\epsfbox{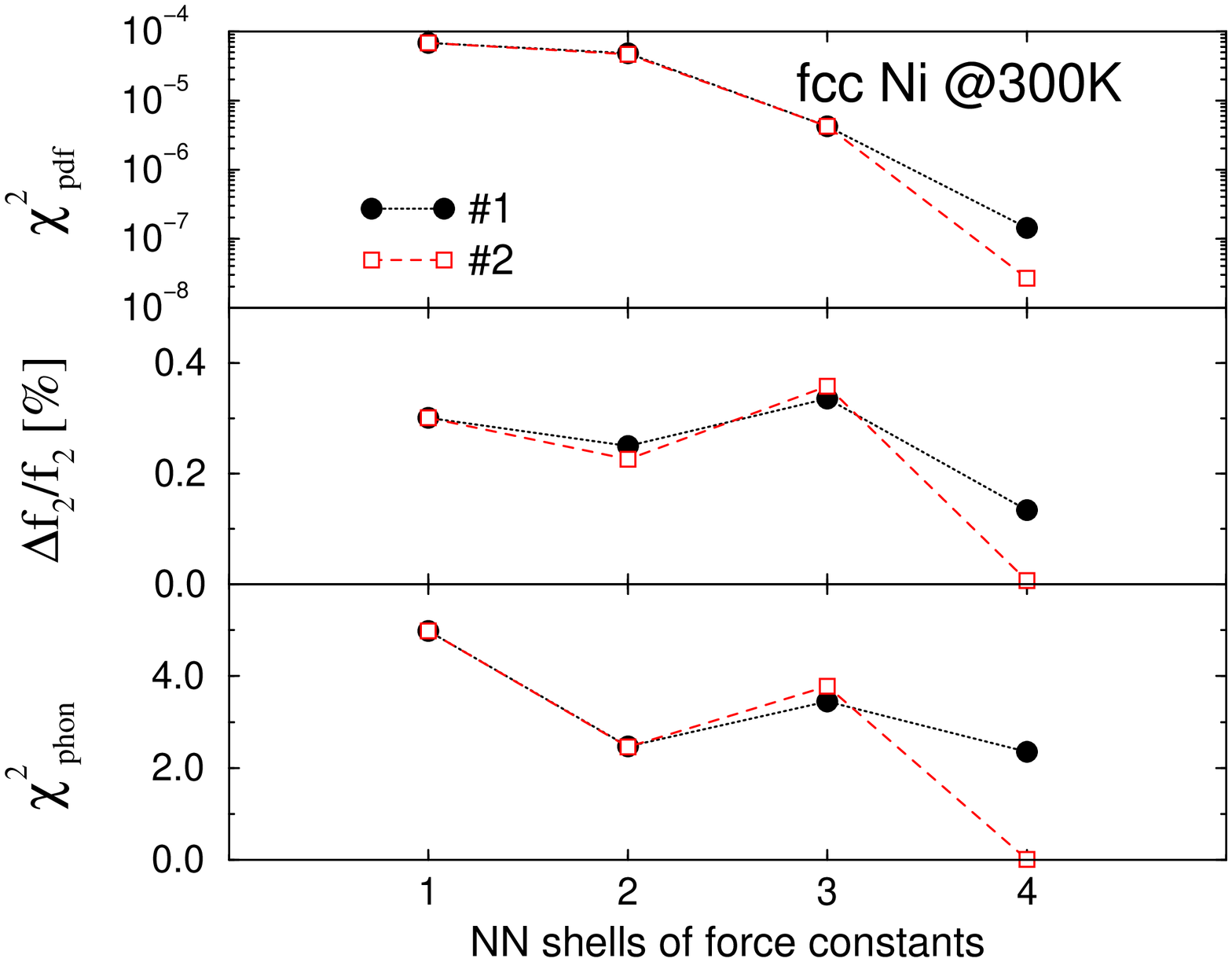}
}
\caption{\label{fig:Ni} 
Fitting the synthetic PDF of Ni, generated with a 4NN BvK force model.
Top: PDF fits vs.\ number of nearest-neighbor shells of force constants for two 
different sets (\#1 and \#2).
Center: Relative error of $f_2$ in percent of 
$f_2^{\rm synth} = 8.03 {\rm THz}$. 
Bottom: Figure of merit of computed phonon dispersions shown in 
Fig.~\ref{Dispersion:Ni}
}
\end{center}
%
\vfill
\noindent
\begin{center}
\epsfysize=80mm
\rotatebox{-90}{
 \epsfbox{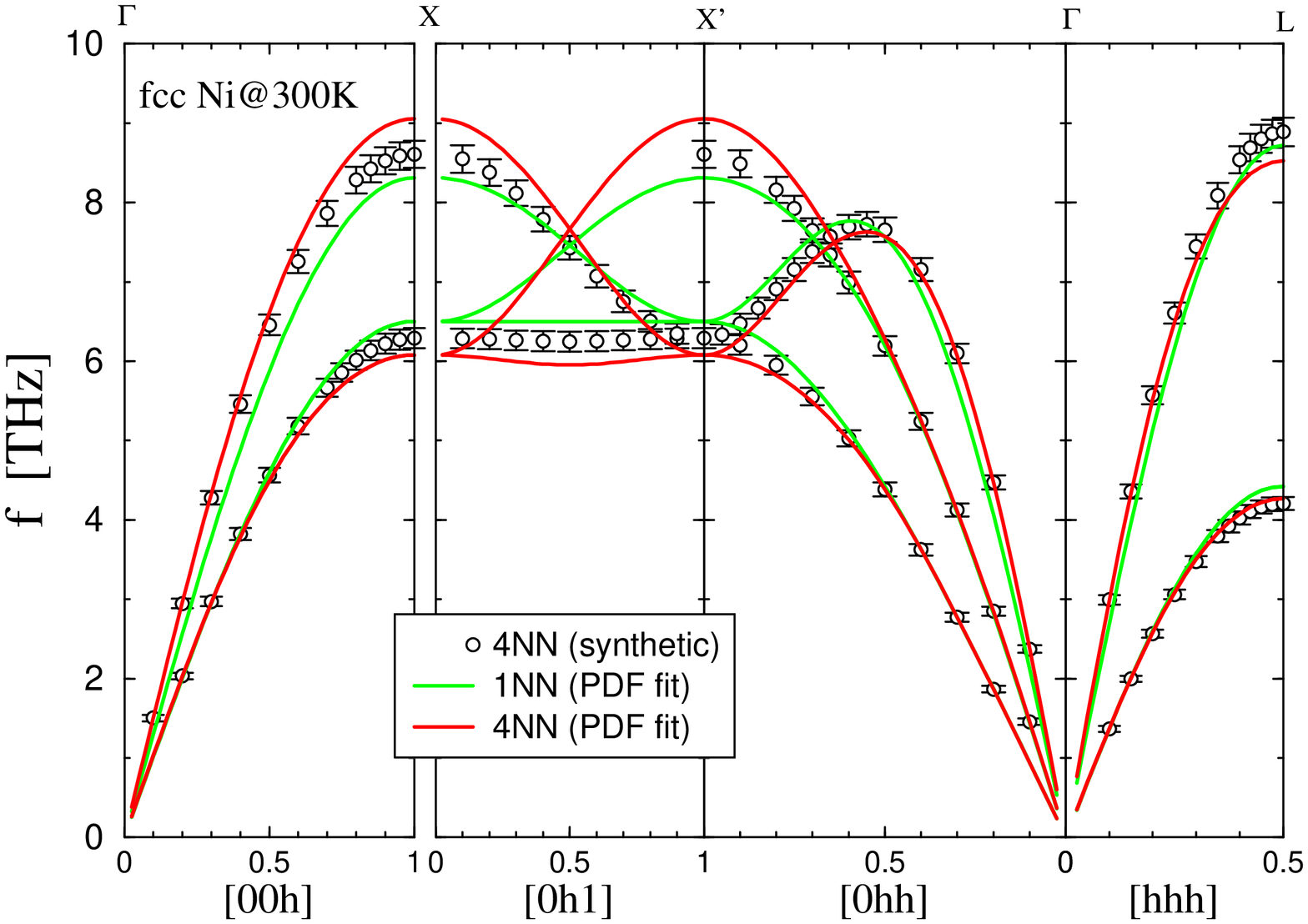}
}
\caption{\label{Dispersion:Ni} (color) Phonon dispersions along high symmetry 
directions of the Brillouin zone obtained from fitting the PDF curve of a 
generalized 4NN BvK force model,\cite{Landolt} using fit models with 
1NN and up to 4NN shells (set \#1).
}
\end{center}
\end{figure}

\begin{figure}[floatfix]
\noindent
\begin{center}
\epsfysize=80mm
\rotatebox{-90}{
\epsfbox{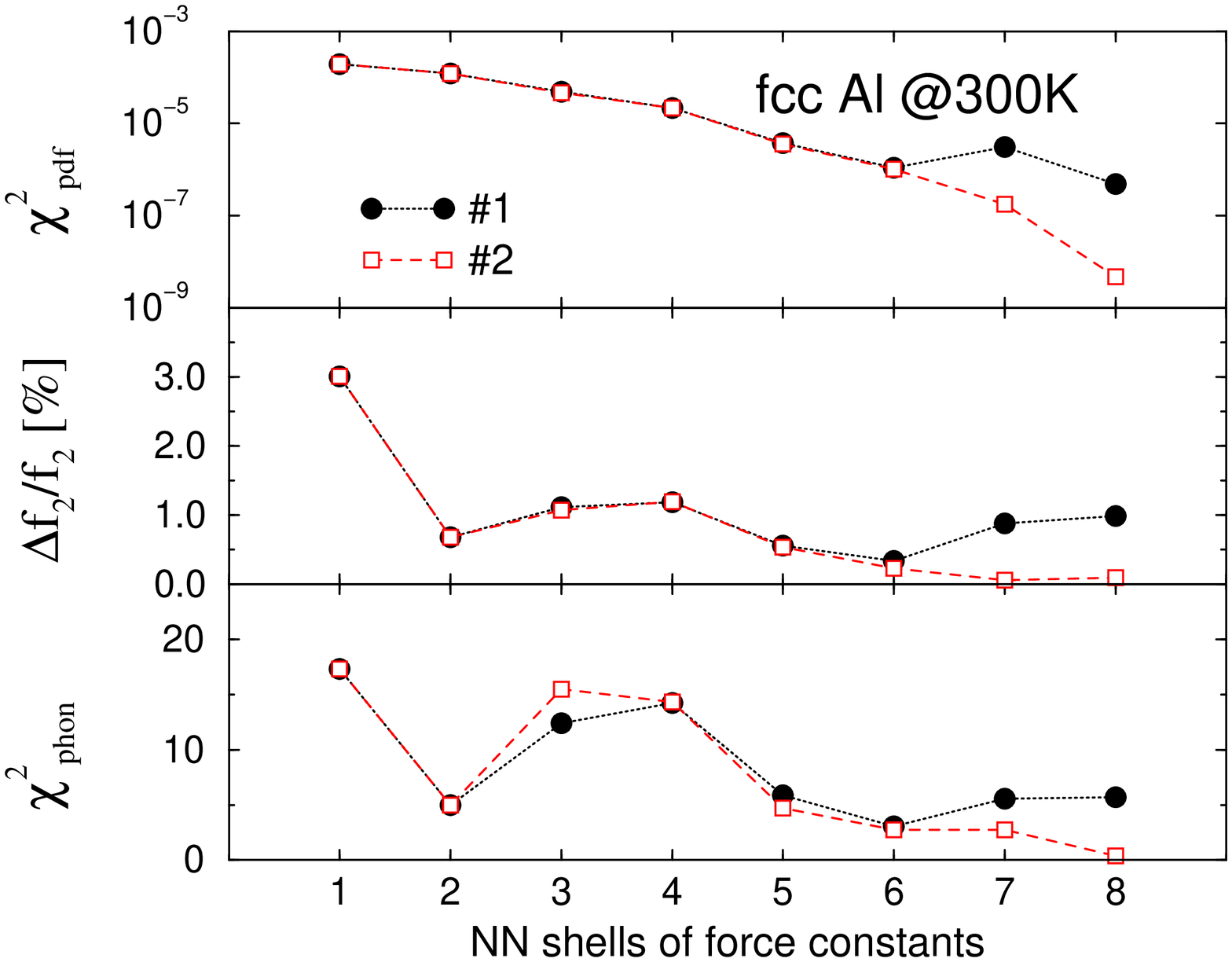}
}
\caption{\label{fig:Al} 
Fitting the synthetic PDF of Al, generated with an 8NN BvK force model.
Top: PDF fits vs.\ number of nearest-neighbor shells of force constants for two 
different sets (\#1 and \#2).
Center: Relative error of $f_2$ in percent of 
$f_2^{\rm synth} = 8.28 {\rm THz}$. 
Bottom: Figure of merit of computed phonon dispersions shown in 
Fig.~\ref{Dispersion:Al}
}
\end{center}
%
\vfill
\noindent
\begin{center}
\epsfysize=80mm
\rotatebox{-90}{
 \epsfbox{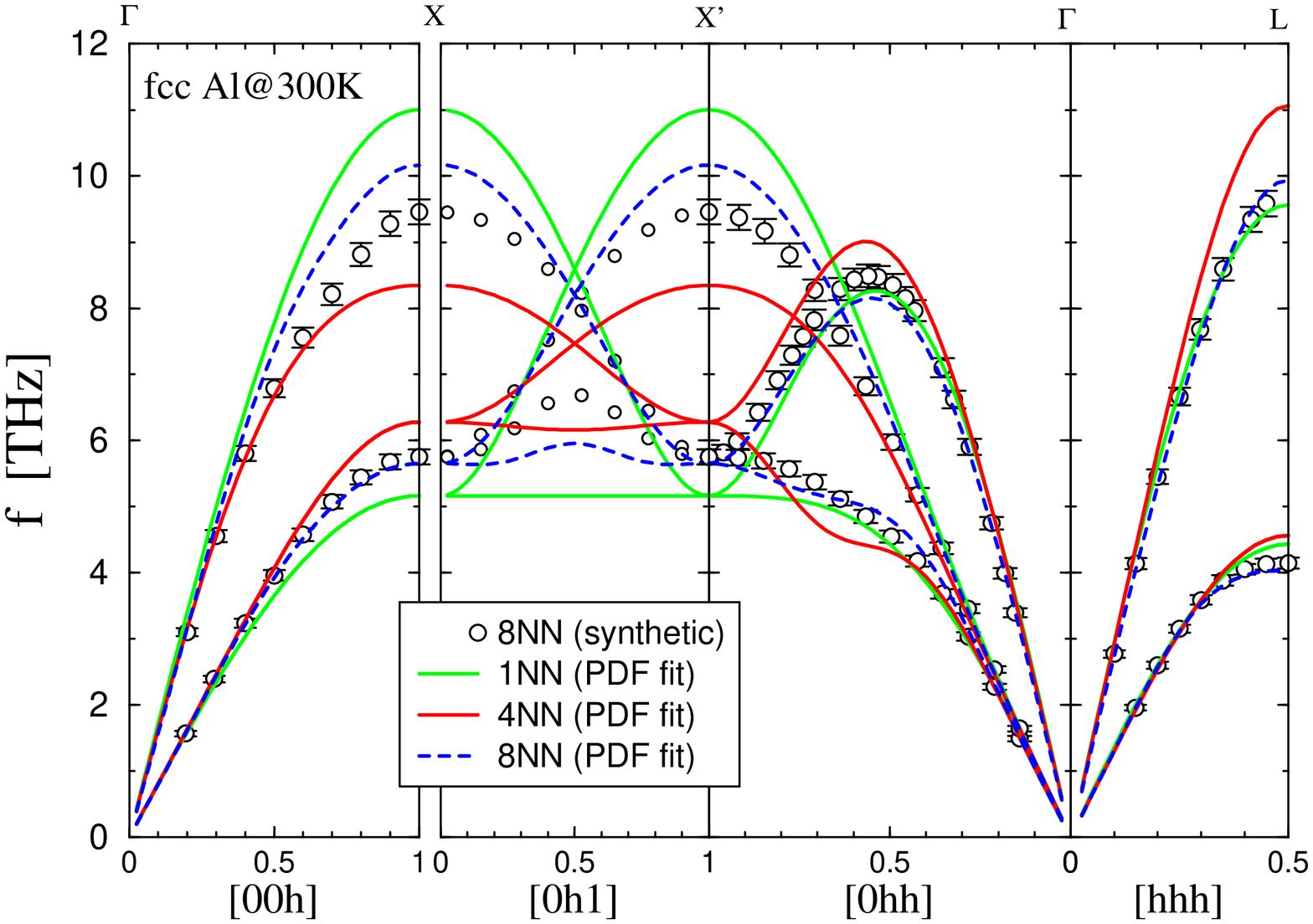}
}
\caption{\label{Dispersion:Al} (color) Phonon dispersions obtained from fitting
the PDF curve of a generalized 8NN BvK force model,\cite{Landolt} 
using fit models with 1NN and up to 4NN and 8NN shells (set \#1).
Only frequencies (circles) with error bars are included in the computation
of $\chi^2_{\rm phon}$ in Fig.~\ref{fig:Al}.
}
\end{center}
\end{figure}

\begin{figure}[floatfix]
\noindent
\begin{center}
\epsfysize=80mm
\rotatebox{-90}{
\epsfbox{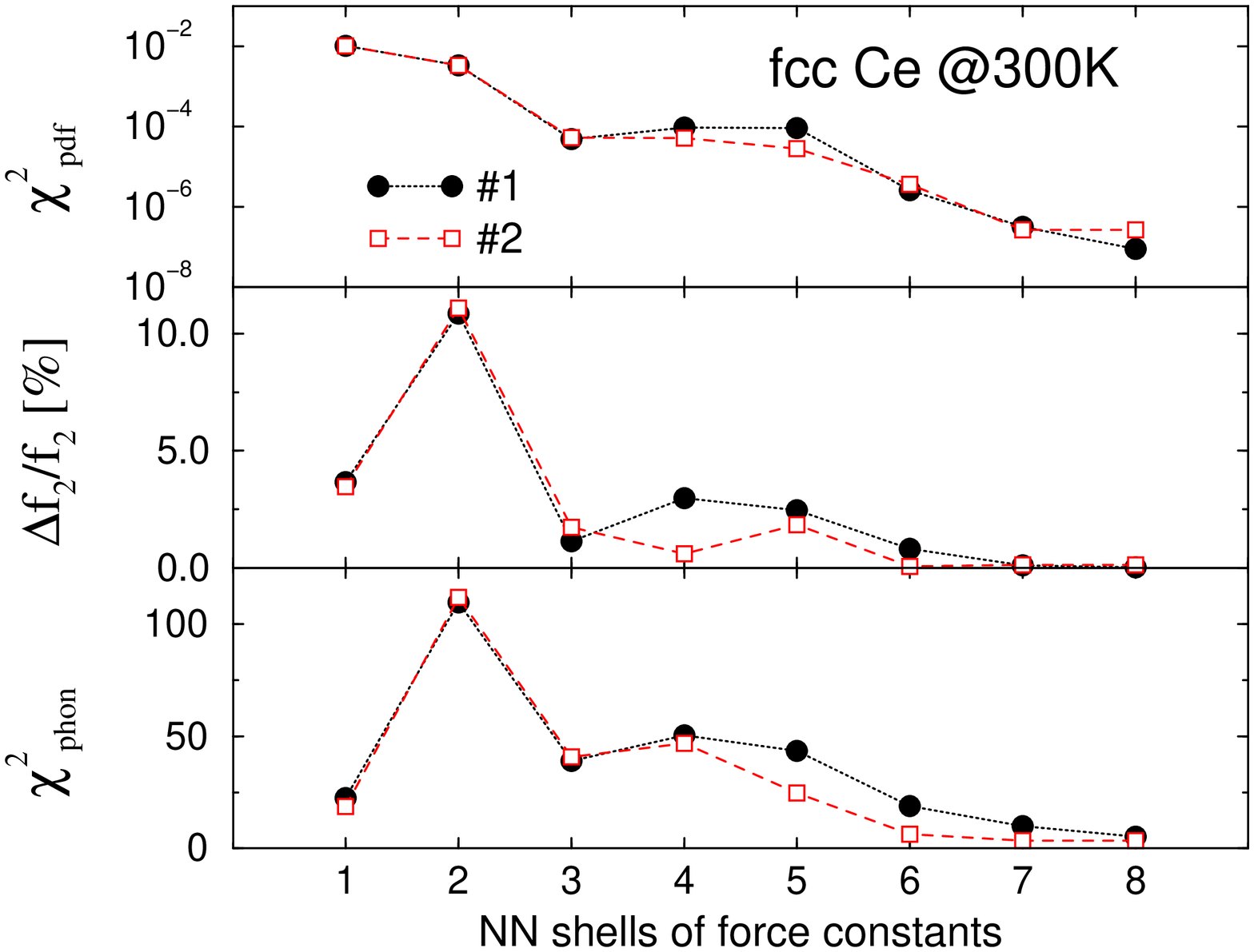}
}
\caption{\label{fig:Ce} 
Fitting the synthetic PDF of Ce, generated with an 8NN BvK force model.
Top: PDF fits vs.\ number of nearest-neighbor shells of force constants for two 
different sets (\#1 and \#2).
Center: Relative error of $f_2$ in percent of 
$f_2^{\rm synth} = 2.48 {\rm THz}$. 
Bottom: Figure of merit of computed phonon dispersions shown in 
Fig.~\ref{Dispersion:Ce}
}
\end{center}
%
\vfill
\noindent
\begin{center}
\epsfysize=80mm
\rotatebox{-90}{
 \epsfbox{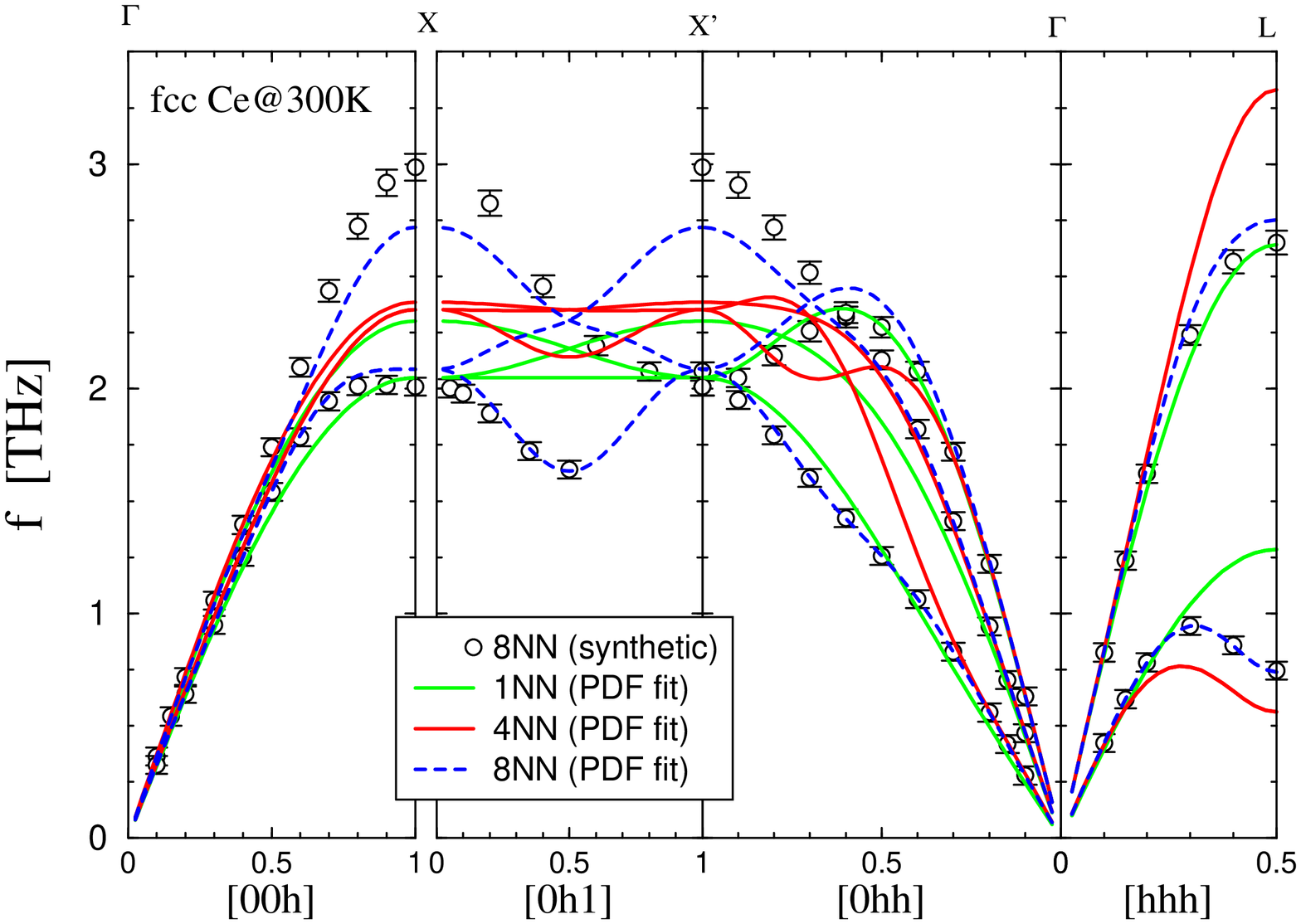}
}
\caption{\label{Dispersion:Ce} (color) Phonon dispersions obtained from fitting
the PDF curve of a generalized 8NN BvK force model,\cite{Landolt} 
using fit models with 1NN and up to 4NN and 8NN shells (set \#1).
}
\end{center}
\end{figure}

\begin{figure}[floatfix]
\noindent
\begin{center}
\epsfysize=80mm
\rotatebox{-90}{
 \epsfbox{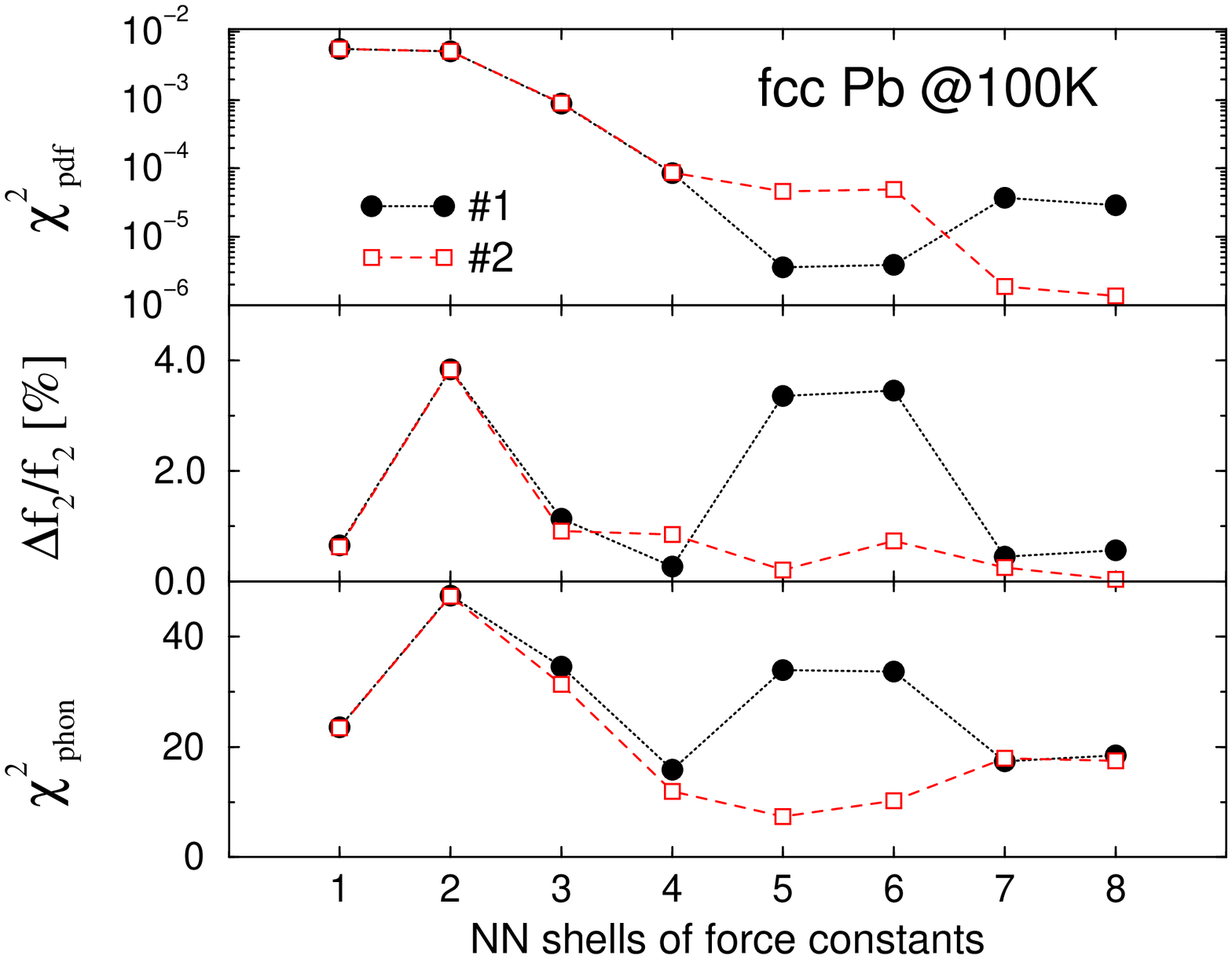}
}
\caption{\label{fig:Pb}
Fitting the synthetic PDF of Pb, generated with an 8NN BvK force model.
Top: PDF fits vs.\ number of nearest-neighbor shells of force constants for two 
different sets (\#1 and \#2).
Center: Relative error of $f_2$ in percent of
$f_2^{\rm synth} = 1.99 {\rm THz}$.
Bottom: Figure of merit of the computed phonon dispersions shown in 
Fig.~\ref{Dispersion:Pb}
}
\end{center}
%
\vfill
\noindent
\begin{center}
\epsfysize=80mm
\rotatebox{-90}{
 \epsfbox{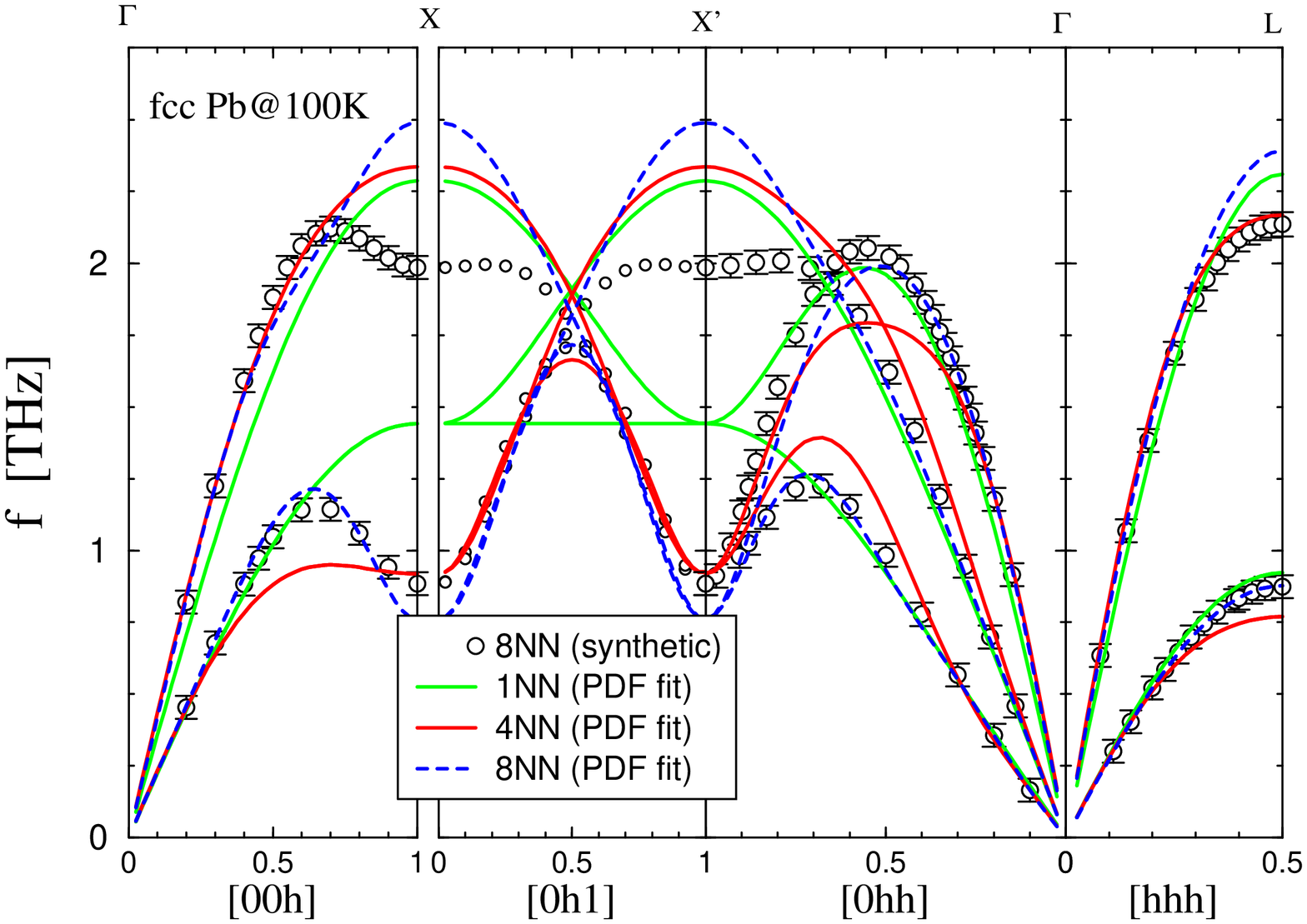}
}
\caption{\label{Dispersion:Pb} (color) Phonon dispersions obtained from fitting
the PDF curve of a generalized 8NN BvK model,\cite{lead} using fit models with
1NN, and up to 4NN and 8NN shells of force constants (set \#1).
Only frequencies (circles) with error bars are included in the computation
of $\chi^2_{\rm phon}$ in Fig.~\ref{fig:Pb}.}
\end{center}
\end{figure}

For each material we used two very different sets of initial force constants  
(set \#1 and set \#2) to start the fit procedure, in order to test the 
robustness of the minimization methods.
In the {\it fcc} simulations shown in 
Figs.~\ref{fig:Ni} through \ref{Dispersion:Pb}, 
we initialized the first shell (1NN) of force constants only 
(i.e., 3 parameters), either by using the correct elastic constants (set \#1),
or by the corresponding values of the 1NN force constants of the synthetic 
model (set \#2).
Note that in cubic systems the three independent elastic constants are
uniquely defined by three force constants (see Appendix).
In the case of the {\it bcc} structures we initialized the combined four 
force constants of the 1NN  and 2NN shells (with the 2NN force constant
$XY_2 \equiv 0$), either by using the correct elastic constants (set \#1),
or by the corresponding values of the synthetic model (set \#2),

Also, for large numbers of force constants the Levenberg-Marquardt
algorithm does not always converge to the global minimum. Instead it gets
easily trapped in local minima, depending on the initial values of the
force constants.
If the initial values are chosen poorly, then this failure is almost
unavoidable. 

\begin{figure}[floatfix]
\noindent
\begin{center}
\epsfysize=80mm
\rotatebox{-90}{
\epsfbox{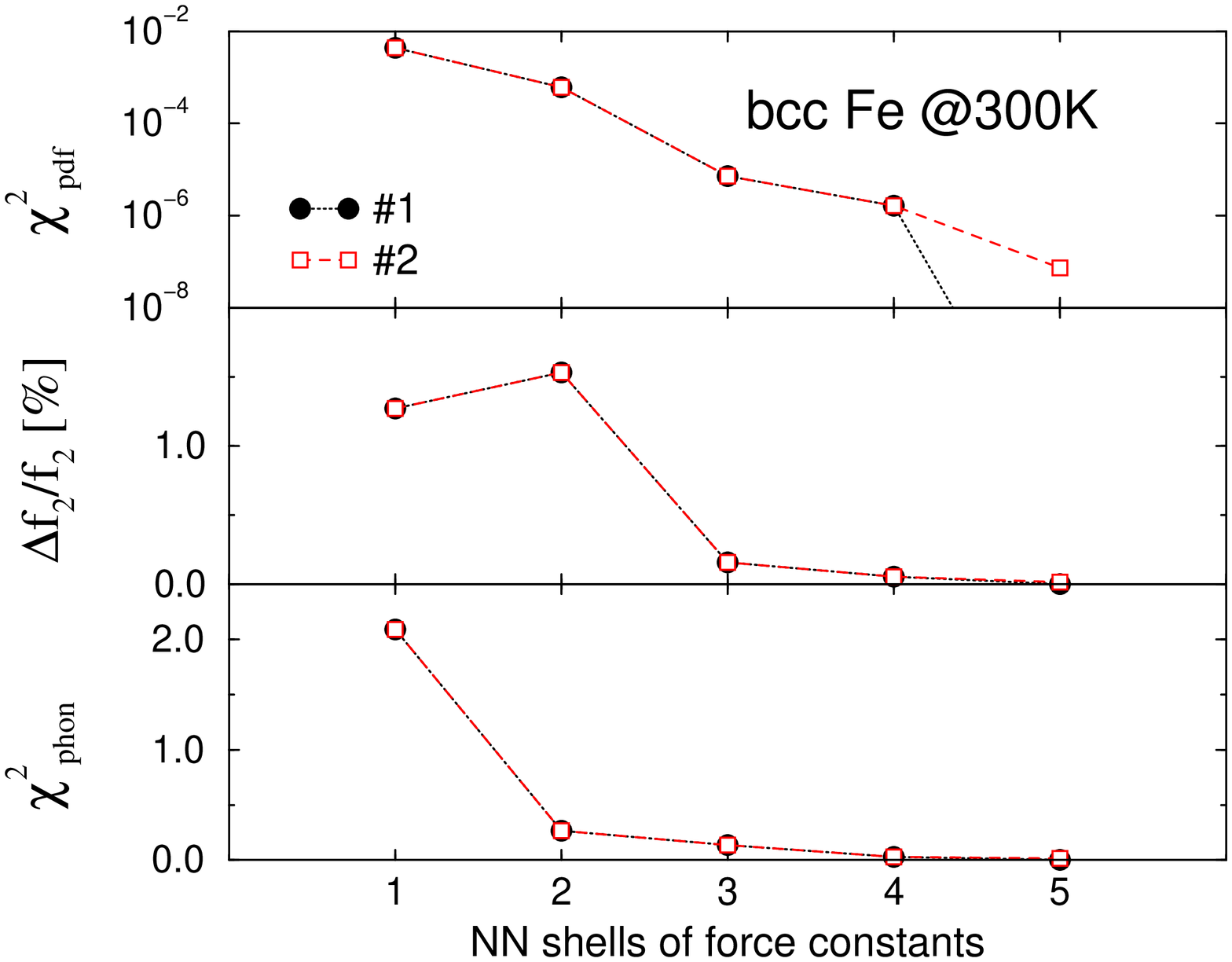}
}
\caption{\label{fig:Fe} 
Fitting the synthetic PDF of Fe, generated with a 5NN BvK force model.
Top: PDF fits vs.\ number of nearest-neighbor shells of force constants for two 
different sets (\#1 and \#2).
Center: Relative error of $f_2$ in percent of 
$f_2^{\rm synth} = 8.77 {\rm THz}$. 
Bottom: Figure of merit of computed phonon dispersions shown in 
Fig.~\ref{Dispersion:Fe}
}
\end{center}
\vfill
\noindent
\begin{center}
\epsfysize=80mm
\rotatebox{-90}{
 \epsfbox{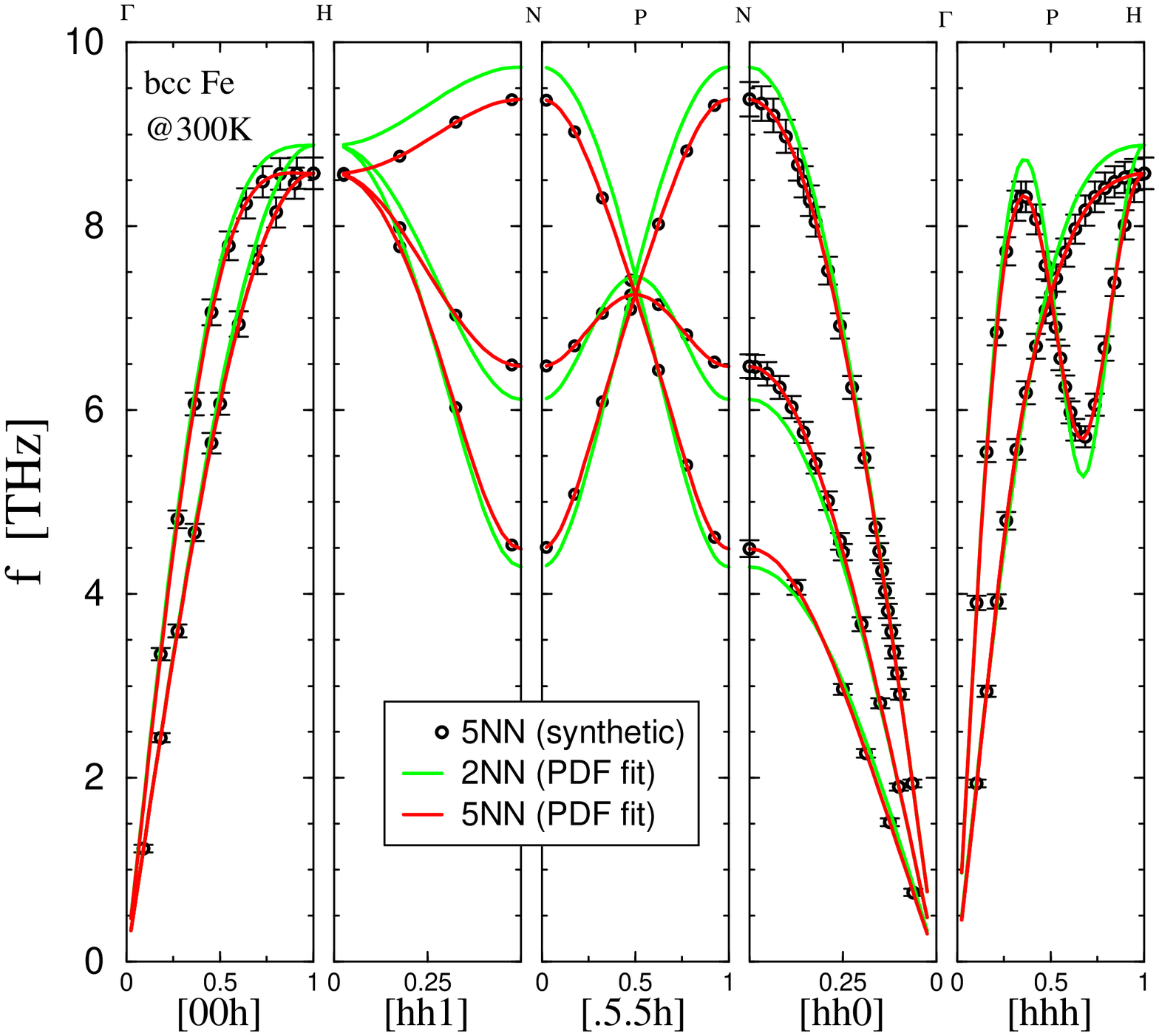}
}
\caption{\label{Dispersion:Fe} (color) Phonon dispersions obtained from fitting
the PDF of a generalized 5NN BvK force model,\cite{Landolt} 
using fit models with up to 2NN and 5NN shells (set \#1).
Only frequencies (circles) with error bars are included in the computation
of $\chi^2_{\rm phon}$ in Fig.~\ref{fig:Fe}.}
\end{center}
\end{figure}

\begin{figure}[floatfix]
\noindent
\begin{center}
\epsfysize=80mm
\rotatebox{-90}{
\epsfbox{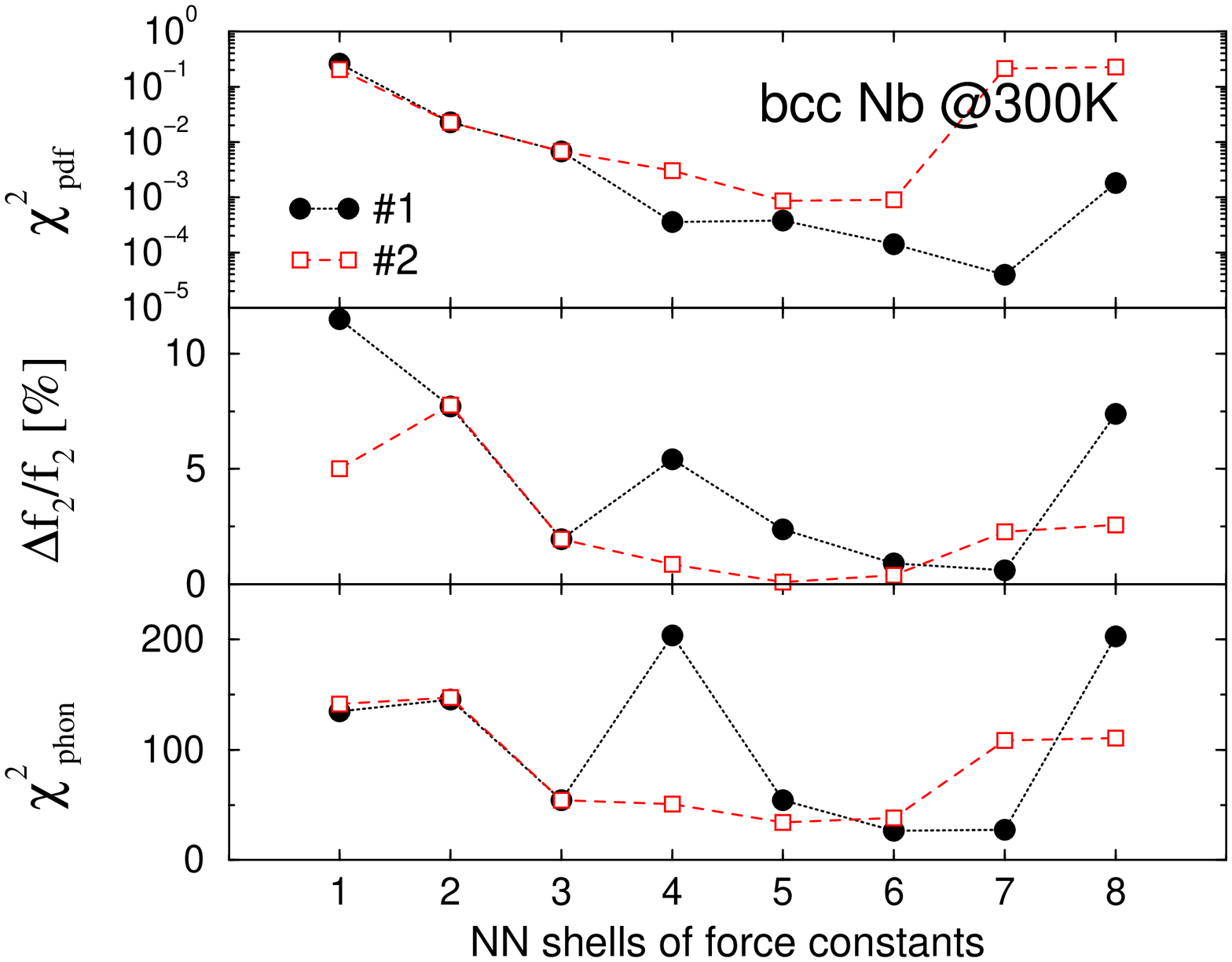}
}
\caption{\label{fig:Nb} 
Fitting the synthetic PDF of Nb, generated with an 8NN BvK force model.
Top: PDF fits vs.\ number of nearest-neighbor shells of force constants for two 
different sets (\#1 and \#2).
Center: Relative error of $f_2$ in percent of 
$f_2^{\rm synth} = 5.86 {\rm THz}$. 
Bottom: Figure of merit of computed phonon dispersions shown in 
Fig.~\ref{Dispersion:Nb}
}
\end{center}
\vfill
\noindent
\begin{center}
\epsfysize=80mm
\rotatebox{-90}{
 \epsfbox{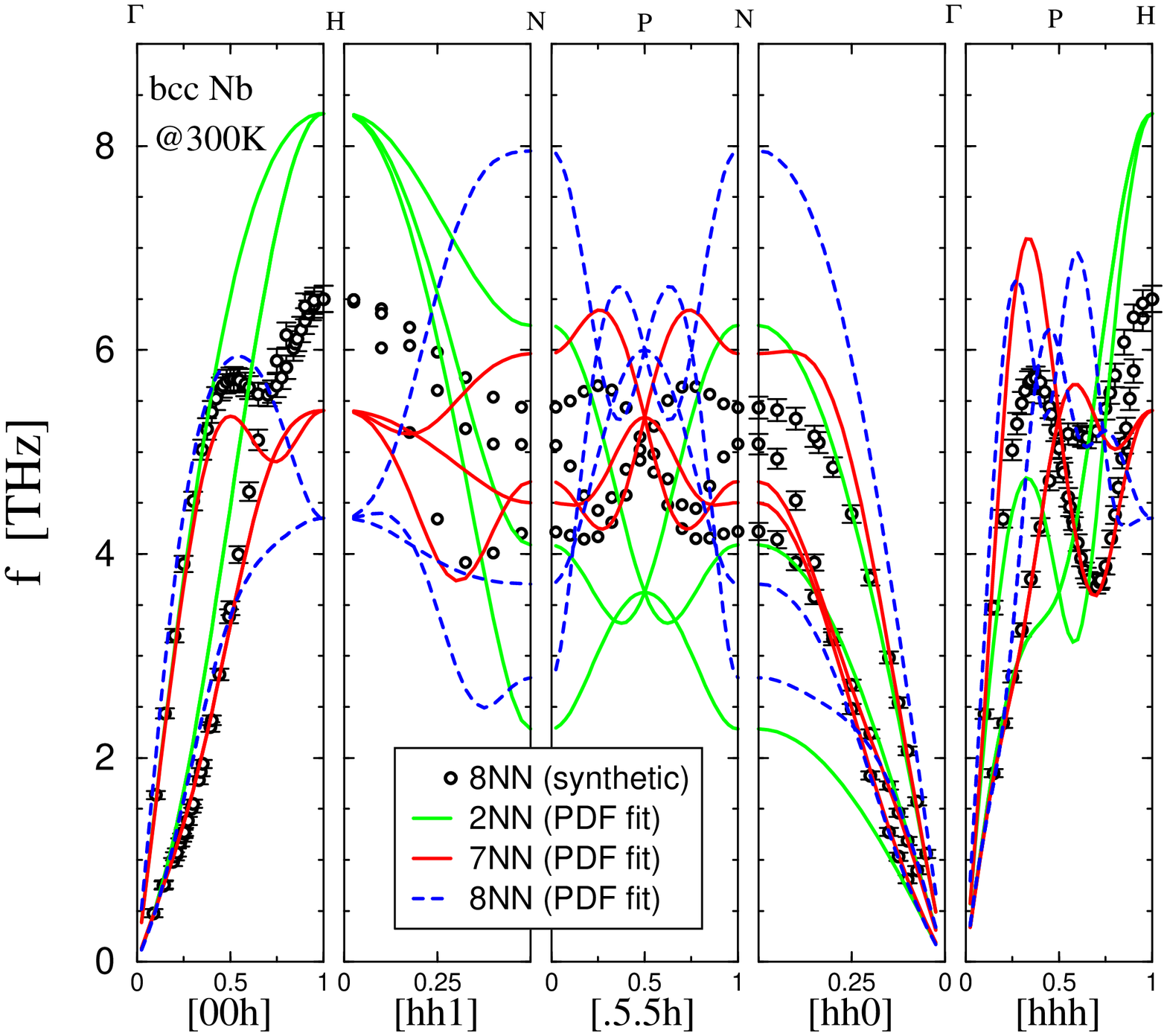}
}
\caption{\label{Dispersion:Nb} (color) Phonon dispersions obtained from fitting
the PDF of a generalized 8NN BvK force model,\cite{Landolt}
using fit models with up to 2NN, 7NN and 8NN shells (set \#1).
Only frequencies (circles) with error bars are included in the computation
of $\chi^2_{\rm phon}$ in Fig.~\ref{fig:Nb}.}
\end{center}
\end{figure}

In the following subsections A through E we address in detail the quality
and difficulties of fitting the PDF curves and investigate the quality of
the corresponding phonon dispersion curves, phonon moments, elastic constants, 
Debye-Waller factors, and PDF peakwidths for a large set of elemental 
materials. We show results for systems that range progressively from very 
simple to very complex phonon models by gradually increasing the number of 
nearest-neighbor interatomic forces in a Born-von K\'arm\'an model.
In figures \ref{fig:Ni} through \ref{Dispersion:Nb} we present a 
comprehensive study of the inverse problem of extracting phonons from
PDF curves from elemental cubic materials.



\subsection{Goodness of Fit}

It is obvious from our results for Ni, Al, Ce and Pb displayed
in Figs.~\ref{fig:Ni} through \ref{Dispersion:Pb}, 
and for Fe and Nb in Figs.~\ref{fig:Fe} through \ref{Dispersion:Nb}, 
that the PDF is rather insensitive to the zone boundary phonons. 
This is best seen for the cases of Al, Ce, Pb and Nb, which have complex 
phonon dispersions with anomalies close to the Brillouin zone boundaries.
In other words, the PDF is not sensitive enough to register the
out-of-phase motions of neighboring atoms.

Also there is no direct correlation between the goodness of the PDF fit
$\chi^2_{\rm pdf}$ and the merit function of the phonon dispersion 
$\chi^2_{\rm phon}$.  For example, in Fig.~\ref{fig:Pb} it can be seen for
the metal Pb that $\chi^2_{\rm pdf}$ of set \#1 is an order of 
magnitude smaller than $\chi^2_{\rm pdf}$ of set \#2 for 4NN and 5NN BvK 
models. However, their corresponding $\chi^2_{\rm phon}$ are reversed. 
This means that a good PDF fit can result in a bad description of the phonon 
dispersion, and vice versa. Similar conclusions follow from our study of Nb in 
Fig.~\ref{fig:Nb}.
This is a very important finding. Hence it is generally not possible to 
provide a useful mapping between PDF spectra and phonon dispersion curves
to solve the inverse problem.
We find that $\chi^2_{\rm pdf}$ and $\chi^2_{\rm phon}$ are correlated overall
only for simple metals (Ni, Ag and Fe) and semi-complex metals (Al and Ce),
and that both fits improve asymptotically by adding more force constants to 
the phonon models.  This demonstrates that our algorithm successfully solves 
the inverse problem for simple cases if the PDF spectra can be obtained with 
arbitrarily high accuracy.
At this point it is not clear if this is a robust result that will survive
once actual \text{experimental} data sets are studied (by including statistical 
and systematic errors into the computation of $S(q)$).
Thus, it is generally not possible to quantify {\it a priori} the quality of
the extracted phonon dispersions based on $\chi^2_{\rm pdf}$.

All PDF fits to the synthetic data sets are almost indistinguishable, as 
follows from the extremely small values of $\chi^2_{\rm pdf}$ (see figures). 
This means that even the crudest BvK phonon model with only 1NN interatomic
forces deviates on average less than $0.1 - 1\%$ from the synthetic PDF data,
which is in good agreement with the analysis of the \textit{forward} problem in
Ref.~\onlinecite{pintschovius01}.
In any realistic experiment, where statistical and systematic measurement
errors may result in a $\chi_{\rm pdf}^2$ of order one, it will be nearly
impossible to attribute a tiny reduction of less than $10^{-3}$ in 
$\chi_{\rm pdf}^2$ to a significantly improved fit. Hence, in a realistic 
simulation a PDF fit with a simple 1NN BvK phonon model will be almost
indistinguishable from one with a more complex 4NN or 8NN BvK model.

\subsection{Phonon Moments}

Since phonon moments are an integrated quantity of the phonon dispersion
curves, it is plausible to expect them to be less sensitive to the details 
of the phonon models being used to generate them. This is indeed the case, 
as can be seen for the second moment shown in figures \ref{fig:Ni}, 
\ref{fig:Al}, \ref{fig:Ce}, \ref{fig:Pb}, \ref{fig:Fe}, and \ref{fig:Nb}. 
Here we focus on the second moment only, because
(1) it is more sensitive to high frequencies (near the zone boundaries)
than the lower moments, and (2) it enters the free energy functional
in the high-temperature limit, and can be obtained independently in a
specific heat measurement.
For simple and semi-complex dispersion curves (Ni, Ag, Fe, and Al, Ce) 
the relative errors of the computed 
second phonon moments track the overall goodness of the PDF fit. 
Unfortunately, this is not true for more complex dispersions (Pb and Nb), 
where we could not establish a correlation between the goodness of 
$\chi^2_{\rm pdf}$ and the relative error of the second moment
$\Delta f_2/f_2^{\rm synth} = |f_2 - f_2^{\rm synth}|/f_2^{\rm synth}$.
However, even systems with complex phonon dispersions allow the
extraction of phonon moments within a few percent of accuracy.
Several years ago Knapp \textit{et al.} \cite{knapp85} arrived at similar 
conclusions while studying the mean-square relative displacement of the 
central atom in \textit{fcc} materials
with extended x-ray absorption fine-structure measurements.

Finally, our computations for Ce, Pb, and Nb show that the second 
phonon moment is too insensitive to the phonons at the zone boundaries
of the Brillouin zone to be useful for determining zone boundary phonons.

\subsection{Elastic Constants}

Dimitrov and co-workers\cite{dimitrov99} suggested that by
adding {\it measured} elastic constants as constraints to the PDF fit 
(constrained fit)
one can improve upon the extracted phonon dispersions. This is contrary to our
own results. We could not observe any significant changes to our extracted
phonon dispersions by using constrained PDF fits; 
\textit{i.e.}, constraining the PDF fit to give 
force constants that yield the correct elastic constants does not result 
in better phonon dispersions near the zone boundaries. Of course it results
in slightly more accurate phonon dispersions near the zone center.
This is not surprising since the elastic constants determine the 
long-wavelength limit of the phonon dispersions at the center of the 
Brillouin zone ($\Gamma$ point) and not at the zone boundaries, where the 
discrepancies between the \textit{synthetic} phonon dispersions and the 
PDF-fitted phonon dispersions are largest. Here the extracted phonon dispersions
could improve by providing additional constraints on zone boundary phonons.

Furthermore, our analysis shows that, in many cases, the unconstrained 
PDF fits already yield elastic constants that deviate only a few percent 
(approximately $1-6\%$) from the values of the elastic constants of the 
\textit{synthetic} data sets, when a significantly 
large number of force constants are being used. By ``sufficiently large'',
we mean at least a 3NN BvK force model for Ni, Ag, Ce and Fe and a 5NN 
BvK model for Al and Pb.
This does not work for Nb, however, where the extracted elastic constants 
are more than $10\%$ off, even for the best fit.
For a definition of the elastic constants in terms of a BvK force model
for \textit{fcc} and \textit{bcc} crystal structures see the Appendix.

\subsection{Debye-Waller Factor}

Reichardt and Pintschovius suggested that it might be possible to improve
the quality of extracted phonon dispersions by adding
constraints to the PDF fit, \textit{e.g.}, thermal parameters independently
measured by a Rietveld analysis.
The thermal parameters, which are given by the exponent of the
Debye-Waller factor, $e^{-2W}$, measure the mean-square atomic displacement
$\langle u^2 \rangle$. For cubic crystals the Debye-Waller exponent 
simplifies to\cite{squires}
\begin{eqnarray}
2 W({\bf q}) &=& \langle ({\bf q}\cdot {\bf u})^2 \rangle 
= q^2 \langle u_{\bf q}^2 \rangle
= \frac{1}{3} q^2 \langle u^2 \rangle  \,,
\end{eqnarray}
where $u_{\bf q}$ is the component of the displacement vector ${\bf u}$ 
in the direction of the scattering vector ${\bf q}$. This result may
be expressed another way,
\begin{eqnarray}
W({\bf q}) &=& \frac{\hbar}{4 M N_{\rm BZ}} \sum_{s}\sum_{{\bf k}\in {\rm BZ}} 
\frac{ \big| {\bf q}\cdot {\bf e}({\bf k}s) \big |^2}{ \omega({\bf k}s) }
\textrm{coth}\left( \frac{\hbar \omega({\bf k}s)}{2 k_B T} \right)
\nonumber\\
&=&
\frac{\hbar q^2}{4 M} \int_{0}^{\infty} d\omega	\ {\cal W}(\omega)
\,,
\end{eqnarray}
with the Debye-Waller spectral function given by
\begin{eqnarray}
{\cal W}(\omega) = \frac{N(\omega)}{\omega} 
\textrm{coth}\left( \frac{\hbar \omega}{2 k_B T} \right)
\,,
\end{eqnarray}
where $M$ is the atomic mass, $\omega({\bf k}s)$ is the angular frequency 
of mode $s$, and ${\bf e}({\bf k}s)$ is its normalized eigenvector.
The phonon density of states is normalized so that
\begin{equation}
\int_{0}^{\infty} d\omega N(\omega)  = 1 \,,
\end{equation}
and $N(\omega) \equiv 0$ for $\omega$ larger than the maximum phonon frequency.
The thermal parameters extracted from our PDF fits are remarkably 
insensitive to the specific form of the phonon model. Even for the
simplest phonon models used, the relative error of $\langle u^2 \rangle$ 
is typically less than $1\%$ (less than $0.1\%$ for Ni, Ag, Al, and Pb) 
and less than $5\%$ for Nb. 
Thus, measured thermal parameters, whose absolute values are known only 
within $5\%$, cannot give improved PDF fits or better phonon dispersions.

\begin{figure}[floatfix]
\noindent
\begin{center}
\epsfysize=75mm
\rotatebox{-90}{
 \epsfbox{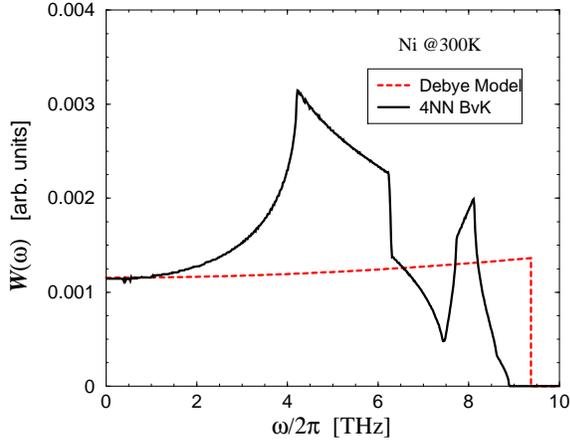}
}
\caption{\label{fig:Debye_Waller_Ni} A comparison of the Debye-Waller
spectral functions of Ni at 300K for a realistic 4NN BvK 
phonon model\cite{Landolt} with its corresponding Debye phonon model.}
\end{center}
\end{figure}

\begin{figure}[floatfix]
\noindent
\begin{center}
\epsfysize=75mm
\rotatebox{-90}{
 \epsfbox{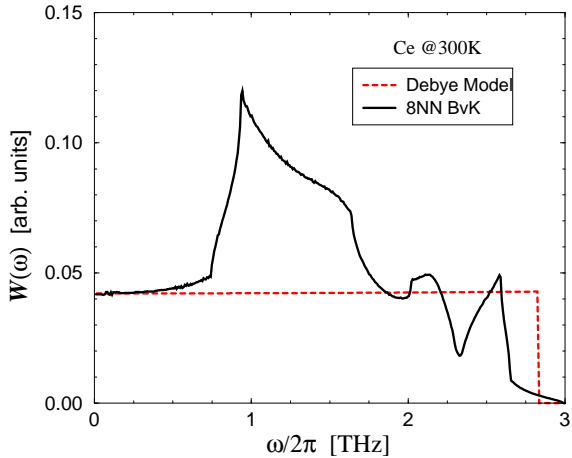}
}
\caption{\label{fig:Debye_Waller_Ce} A comparison of the Debye-Waller
spectral functions of Ce at 300K for a realistic 8NN BvK 
phonon model\cite{Landolt} with its corresponding Debye phonon model.}
\end{center}
\end{figure}

At this point a caveat is needed for using elastic constants
(derived from long wavelength modes) and a Debye phonon model to estimate
the Debye-Waller factor, or when comparing Debye temperatures
with Debye-Waller factors measured at high temperatures. 
Figs.~\ref{fig:Debye_Waller_Ni} and \ref{fig:Debye_Waller_Ce} 
show very clearly for Ni and Ce that at room temperature
the Debye-Waller spectral function ${\cal W}(\omega)$
estimated from the long wavelength modes, is only a crude 
approximation. It consistently underestimates the contribution of
the intermediate frequency region (transverse modes near the zone boundary), 
and describes the high frequency region (longitudinal modes near the zone
boundary) only on average.
Here the Debye-Waller exponents $2W$ computed from 
a Debye phonon model are $20\%$ too small, compared to the ones
using more realistic lattice dynamical models.
The other materials studied show similar discrepancies, except in the case
of Nb where both the Debye and lattice model calculations
of the thermal parameters accidentally agree within $1$ percent.
Indeed, it is well known that the Debye-Waller factor is rather insensitive
to the detailed form of the phonon dispersion, and at high temperatures it
depends only on a single parameter, namely, the inverse-squared phonon 
moment.\cite{marshall71}

\begin{figure}[floatfix]
\noindent
\begin{center}
\epsfysize=75mm
\rotatebox{-90}{
 \epsfbox{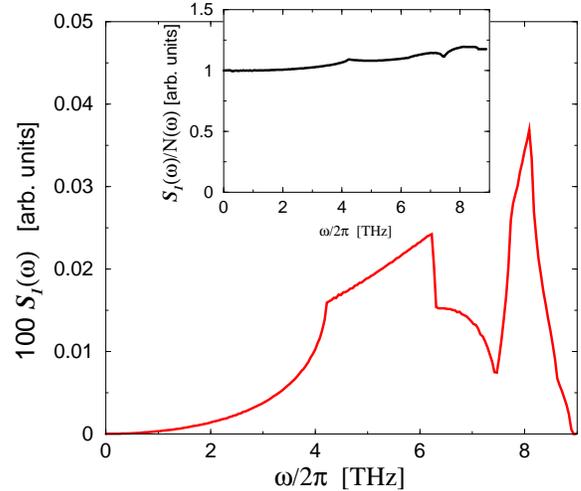}
}
\caption{\label{fig:peakwidth_Ni} The spectral function of the first PDF
peak of Ni at 300K. Insert: ${\cal S}_1(\omega)/N(\omega)$ is nearly 
constant for frequencies $\omega < \omega_{\rm max}/2$. This frequency
range is significantly smaller for Pb and Nb and marks the deviation
from a Debye-model phonon spectrum.
}
\end{center}
\end{figure}

\begin{figure}[floatfix]
\noindent
\begin{center}
\epsfysize=75mm
\rotatebox{-90}{
 \epsfbox{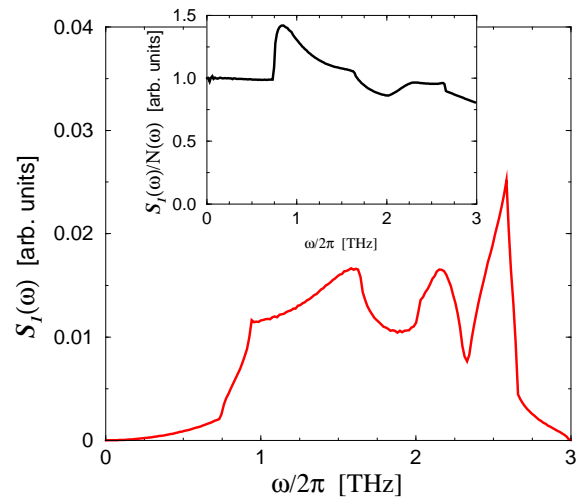}
}
\caption{\label{fig:peakwidth_Ce} The spectral function of the first PDF
peak of Ce at 300K. Insert: ${\cal S}_1(\omega)/N(\omega)$ is nearly 
constant for frequencies $\omega < \omega_{\rm max}/4$.
}
\end{center}
\end{figure}

\subsection{PDF Peakwidths}

For studying local atomic structure properties in semiconductors 
Chung and Thorpe \cite{chung97} used a real space approach
to compute the PDF. They showed in the harmonic approximation that
the PDF is approximately a series of Gaussian peaks, each centered at
distance $r_i$ with width $\sigma_i$,
\begin{eqnarray}\label{peakwidth}
\sigma^2_i &=& \frac{\hbar}{M N_{\rm BZ}}
\sum_s \sum_{{\bf k} \in {\rm BZ}} {\rm coth}
 \left( \frac{\hbar \omega({\bf k}s)}{2 k_B T} \right)
 \frac{| \hat{r}_i \cdot {\bf e}({\bf k}s) |^2}{\omega({\bf k}s)}
\nonumber \\ && \hspace*{20mm} \times
 \big[ 
   1 - \cos( {\bf k} \cdot {\bf r}_i)
 \big]
\,.
\end{eqnarray}
Here ${\bf r}_i$ is a position vector of an atom in the $i$-th NN shell 
measured relative to an atom at the origin, and $\hat{r}_i = {\bf r}_i/r_i$.
In the limit ${\bf k} \cdot {\bf r}_i \gg 1$ the cosine term
oscillates so rapidly across the BZ that its average vanishes and the 
peakwidths of far-out atoms contain the same information as the Debye-Waller
exponent, namely,
$\lim_{i\to \infty}\sigma^2_i \equiv \sigma^2_\infty = 4W(q)/q^2$.
This explains why the PDF fits reproduce the Debye-Waller factors so well,
and that adding thermal parameters cannot provide any extra constraints.

On the other hand, the width of the first PDF peak (which is the sum of the
auto-correlation and cross-correlation parts of the 
displacement-displacement function) exhibits the largest deviations
from the long-range value $\sigma^2_\infty$ (which is purely the 
auto-correlation part of the displacement-displacement function), 
due to the short-range nature of correlations of lattice vibrations.  Note 
that in the real space approach all multiphonon processes have been included.
Expanding $\cos({\bf k} \cdot {\bf r}_i)$ in Eq.~\ref{peakwidth}, it follows
that at high temperatures and for low frequencies the spectral function
of $\sigma^2_i$ is, in first-order approximation, dominated by the phonon 
density of states. We define the spectral function
${\cal S}_i(\omega)$ of the $i$-th peak as 
\begin{eqnarray}
\sigma^2_i &=& \frac{\hbar}{M}
\int_0^\infty d\omega \ {\cal S}_i(\omega)
\,, 
\\
{\cal S}_i(\omega) &=& 
{\cal W}(\omega)
\left\langle |\hat{r}_i \cdot {\bf e}({\bf k}s)|^2 
  [1 - \cos({\bf k} \cdot {\bf r}_i)] 
\right\rangle_{\omega}
,
\end{eqnarray}
where $\langle \dots \rangle_{\omega}$
is a normalized average over phonon modes and {\bf k}-points at fixed 
frequency, $\omega({\bf k}s) = \omega$, 
and ${\cal S}_i(\omega) \equiv 0$ for frequencies
larger than the maximum phonon frequency. For sufficiently small frequencies
we get
$\left\langle |\hat{r}_1 \cdot {\bf e}({\bf k}s)|^2 [1 - \cos({\bf k} \cdot
{\bf r}_1)] \right\rangle_{\omega} \sim k^2 a^2 \sim (a/c)^2 \omega^2$,
with the average sound speed $c$ and lattice constant $a$.
For temperatures $T \gg \hbar\omega/2 k_B$, the width of the first peak is a 
rough measure of the integrated phonon density of states times 
$\frac{\hbar}{M}\frac{a^2}{c^2}$, as can be seen in the inserts of
Figs.~\ref{fig:peakwidth_Ni} and \ref{fig:peakwidth_Ce}.

It is not too surprising to find that the widths of the first peak and that 
of very distant peaks, or equivalently the Debye-Waller factor, are
crudely approximated by long wavelength phonons, although for quite different 
physical reasons. 
Since for purely elastic neutron scattering the peakwidths
are identical, $\sigma^2_i = \sigma^2_\infty = 4W(q)/q^2$,\cite{marshall71} 
and in most metals inelastic scattering (one-phonon, two-phonon, and 
higher-order processes) leads only to small corrections $\triangle\sigma_1$
of order $10-30\%$ in the width of the first peak $\sigma_1$, 
one does not expect large variations in $\sigma_i$. Hence the 
magnitude of the peakwidths will be approximated in leading order
by the long wavelength phonons.
For example, in the case of Ni we find a difference of 
$\triangle\sigma_1 / \sigma_1 \approx 13\%$ between the fitted PDF 
curve that includes only elastic processes versus elastic plus 
one-phonon processes.
For Al and Pb we obtain $\triangle\sigma_1 / \sigma_1 \approx 25\%$.
These one-phonon corrections are largest for the first PDF peak 
and much smaller for the other peaks.\cite{earlier,pintschovius01}

\section{Conclusions}

In conclusion, our study shows that one cannot obtain accurate phonon 
dispersions from an inverse analysis of the pair-density function, unless the 
lattice dynamics are simple and fully described by a few phonon 
parameters, as in the cases of {\it fcc} Ni and Ag and {\it bcc} Fe,
for example.  A semi-quantitative picture 
of the phonon dispersion may be obtained in simple and semi-complex
metals, but not in metals with complex phonon dispersions. 
In principle, in simple metals phonon frequencies can be extracted within
a few percent accuracy ($\sim 2-8\%$) in the entire Brillouin zone, 
whereas in semi-complex and complex metals such accuracy applies only to a small
fraction of the Brillouin zone centered around the $\Gamma$-point with
wave vector $k \lesssim \pi/4 a$.
We found numerically that the pair-density function provides an overall 
account of the lattice dynamics by yielding phonon moments within 
a few percent accuracy.  
In other words, a rather simple phonon model suffices to describe the 
dynamics embedded in powder diffraction data.
Neutron or x-ray PDF studies play an important role in the studies
of the local structure of crystals, but cannot provide deeper insight 
into the dynamics of lattice vibrations.
A more promising approach for extracting phonons from powders or polycrystals
may be the analysis of time-of-flight spectra from inelastic neutron
scattering.\cite{buchenau83,heiroth86}

\begin{acknowledgments}
We thank D. Wallace, A. Lawson, H. R\"oder, L. Pintschovius,
D. Strauch and G. Eckold for many helpful discussions and especially
J. Wills for helping with the symmetry operations and {\bf k}-point summations.
This work was supported by the Los Alamos National Laboratory under the 
auspices of the U.S.~Department of Energy.
\end{acknowledgments}

\appendix*
\section{Elastic Constants in a Born-von K\'arm\'an force model}

In a cubic crystal there are only three independent elastic 
constants.\cite{wallace}
For \textit{fcc} crystal structures with lattice constant $a$ these are 
related to the force constants (up through 8NN interatomic shells) by
\begin{eqnarray}
&a C_{11} = 4 XX_1 + 4 XX_2 + 16 XX_3 + 8 YY_3 +  16 XX_4 + 
&\nonumber\\ &
 36 XX_5 + 4 YY_5 + 16 XX_6 + 72 XX_7 + 
&\nonumber\\ &
 32 YY_7 + 8 ZZ_7 + 16 XX_8	\,,
\end{eqnarray}
\begin{eqnarray}
&a C_{44} = 2 XX_1 + 2 ZZ_1 + 4 YY_2 + 4 XX_3 + 20  YY_3 + 
&\nonumber\\ &
 8 XX_4 + 8 ZZ_4 + 2 XX_5 + 18 YY_5 + 20 ZZ_5 + 
&\nonumber\\ &
 16 XX_6 + 20 XX_7 + + 40 YY_7  + 
&\nonumber\\ &
52 ZZ_7 + 16 YY_8 \,,
&
\end{eqnarray}
\begin{eqnarray}
&a (C_{12}+C_{44}) = 4 XY_1 + 8 YZ_3 + 32 XZ_3 + 16 XY_4 + 
&\nonumber\\ &
 24 XY_5 + 32 YZ_6 + 96 XY_7 + 
&\nonumber\\ &
48 XZ_7 + 32 YZ_7
\,,
&
\end{eqnarray}
and for \textit{bcc} structures these are given by
\begin{eqnarray}
&a C_{11} = 2 XX_1 + 2 XX_2 + 8 XX_3 + 18 XX_4 + 4 YY_4 + 
&\nonumber\\ &
 8 XX_5 + 8 XX_6 + 36 XX_7 + 2 ZZ_7 + 
&\nonumber\\ &
32 XX_8 + 8 YY_8
\,,
&
\end{eqnarray}
\begin{eqnarray}
&a C_{44} = 2 XX_1 + 2 YY_2 + 4 XX_3 + 4 ZZ_3 + 2 XX_4 + 
&\nonumber\\ &
 20 YY_4 + 8 XX_5 + 8 YY_6 +
 20 XX_7 + 
& \nonumber\\ &
 18 ZZ_7 + 4 XX_8 + 16 YY_8 + 20 ZZ_8
\,,
&
\end{eqnarray}
\begin{eqnarray}
&a (C_{12} + C_{44} ) = 4 XY_1 + 8 XY_3 + 4 YZ_4 + 24 XY_4 + 
&\nonumber\\ &
 16 XY_5 + 24 YZ_7 + 36 XY_7 + 32 XY_8
\,,
&
\end{eqnarray}
where we followed the derivation of Squires.\cite{squires63,squires62}
Our elastic constants agree with those in Ref.~\onlinecite{squires63}
where comparison is possible, except for the term $4 YY_4$ in $C_{11}$
for \textit{bcc} structures.
The generalized BvK force matrix of the $n$-th NN interatomic
shell is defined by
\begin{equation}
\bm{\Phi}_{\rm nNN} = 
\left(
 \begin{tabular}{ccc}
  $XX_n$	&	$XY_n$	&	$XZ_n$	\\
  $XY_n$	&	$YY_n$	&	$YZ_n$	\\
  $XZ_n$	&	$YZ_n$	&	$ZZ_n$
 \end{tabular}
\right)
\,.
\end{equation}
Its symmetry properties are listed in Table~\ref{table:FCM}.

\begin{table}[floatfix]
\caption{\label{table:FCM}Symmetries of the generalized BvK force constant 
matrix $\bm{\Phi}_{\rm nNN}$ for monatomic \textit{fcc} and \textit{bcc} 
crystal structures. The lattice indexes $[h_1 h_2 h_3]$ refer to lattice 
positions $(h_1, h_2, h_3) a/2$ with $h_1 \geq h_2 \geq h_3$. }
\begin{ruledtabular}
\begin{tabular}{ccc}
shell	&	fcc		&	bcc	\\
\hline
1NN	&	[110]		&	[111]	\\
	& $XX_1=YY_1$		& $XX_1=YY_1=ZZ_1$	\\
	& $XZ_1=YZ_1=0$		& $XY_1=XZ_1=YZ_1$	\\
2NN	&	[200]		&	[200]		\\
	& $YY_2=ZZ_2$		& $YY_2=ZZ_2$		\\
	& $XY_2=XZ_2=YZ_2=0$	& $XY_2=XZ_2=YZ_2=0$	\\
3NN	&	[211]		& 	[220]		\\
	& $YY_3=ZZ_3$		& $XX_3=YY_3$		\\
	& $XY_3=XZ_3$		& $XZ_3=YZ_3=0$		\\
4NN	&	[220]		& 	[311]		\\
	& $XX_4=YY_4$		& $YY_4=ZZ_4$		\\
	& $XZ_4=YZ_4=0$		& $XY_4=XZ_4$		\\
5NN	& 	[310]		&	[222]		\\
	& $XZ_5=YZ_5=0$		& $XX_5=YY_5=ZZ_5$	\\
	&	--		& $XY_5=XZ_5=YZ_5$	\\
6NN	&	[222]		&	[400]		\\
	& $XX_6=YY_6=ZZ_6$	& $YY_6=ZZ_6$		\\
	& $XY_6=XZ_6=YZ_6$	& $XY_6=XZ_6=YZ_6=0$	\\
7NN	&	[321]		&	[331]		\\
	& 	--		& $XX_7=YY_7$, $XZ_7=YZ_7$ \\
8NN	&	[400]		&	[420]		\\
	& $YY_8=ZZ_8$		& $XZ_8=YZ_8=0$		\\
	& $XY_8=XZ_8=YZ_8=0$	&	--		\\
\end{tabular}
\end{ruledtabular}
\end{table}


\begin{thebibliography}{10}
\bibitem{multiphonon}
We mimick the effects of multiphonon processes by adding a smooth function,
$S_m(q) = 1-\exp(-2W(q)/2.4)$, to the computation of the coherent static
structure factor $S(q) = S_0(q) + S_1(q) + S_m(q)$. In addition to
the elastic ($S_0$) and 1-phonon ($S_1$) coherent scattering contributions,
$S_m$ guarantees the correct asymptotic limits of $S(q)$, namely,
$\lim_{q \to \infty} S(q) = 1$ and 
$\lim_{q \to 0} S(q) = 0$.
\bibitem{lead}
We used a slightly improved set of force constants for Pb that results in
more accurate phonon frequencies near the zone boundaries than the one in 
Ref.~\onlinecite{Landolt}.
\end{thebibliography}
\end{document}